# Evidence of thermal effects in high power $Er^{3+}$-$Yb^{3+}$ fiber laser

**G. Canat, J.-C. Mollier**

*DOTA, ONERA, Chemin de la Hunière, 91761 Palaiseau CEDEX France,*
*guillaume.canat@onecert.fr*

**Y. Jaouën**

*GET Telecom Paris, CNRS UMR 5141, 46 rue Barrault, 75634 Paris, France*

**B. Dussardier**

*LPMC, CNRS, URA 190, Université de Nice, Parc Valrose 06108, Nice Cedex 2, France*

We analyse the influence of heat generation by non radiative transitions in high power 1.55µm double cladding Erbium-Ytterbium fiber laser. At strong pumping rates, 1µm lasing can start due to parasitic reflections. We present a model including heat generation and its effect on the Stark level population using the MacCumber relation. Heat generation plays then a significant role and improves the 1.5µm laser efficiency by increasing the 1µm threshold.

Double-cladding fiber lasers are appealing sources due to their high brightness, high efficiency and versatility. They address many potential applications in communication, remote sensing and imaging. High output power can be obtained using double-clad doped fiber pumped with broad area high power pump laser diodes. Double-clad fibers with pure $Yb^{3+}$ doping or $Er^{3+}/Yb^{3+}$ co-doping can be used to build efficient lasers at 1 μm range and 1.55 μm range respectively: 103 W at 1.5 μm [1] and 1 kW at 1 μm [2] have been demonstrated in slightly multimode fibers. Several hundreds of watts to several kilowatts of pump power have to be absorbed. Due to the quantum defect between pump and laser photons, heat is generated into the active medium and thermal effects start to be significant in these lasers. Fibers take advantage of a much higher ratio volume / cross section than rods and disks frequently used in solid state laser. Heat generation in kilowatt Ytterbium fiber lasers has been previously studied [3]. The quantum efficiency of Yb fiber laser is closed to 90% and these lasers were generally considered immune to thermal effects up to several kilowatts pump power [4]. The theoretical efficiency of Er/Yb lasers is limited to 63 % but the actual efficiency typically between 30 % and 50%. It follows that a much larger amount of power is converted into heat. To the best of our knowledge, no work as yet been done on the effect of self induced heating in Erbium-Ytterbium codoped fiber laser. In spite of the large dissipation power of fibers, the core temperature can rise by several hundred of degrees. In this paper, we study the influence of self-induced heating due to pump absorption and its effect on lasing efficiency in $Er^{3+}/Yb^{3+}$ co-doped fiber laser up to 100 W output power.

(note: pour éviter des confusions entre différents codopants (cas de yb et Er) et aider le lecteur, on note parfois les niveaux comme ceci: "Er: $^4I_{13/2}$")

The Er/Yb system in silica involves absorption of pump photons between the Ytterbium manifolds $^2F_{7/2}$ (fundamental) and $^2F_{5/2}$ (excited). The energy is then resonantly transferred to the Erbium $^4I_{11/2}$ level which desexcites to $^4I_{13/2}$ level. The lasing transition takes place between $^4I_{13/2}$ and $^4I_{15/2}$ levels. Due to the glass field, all these manifolds are split by Stark effect into sublevels (fig. 1). We have measured Er and Yb absorption and fluorescence on specially prepared singlemode phospho-alumino-silicate Er/Yb doped fibers. Even though the tail exhibits inhomogeneous broadening, the secondary emission peak is clearly visible at 1020 nm (fig. 2). The emission beyond this peak is much larger than absorption when the Yb inversion is high, the net gain beyond 1030 nm is very high. At very high pump power, the residual reflections can thus lead to 1.06 µm parasitic lasing (plutôt 1015à 1025 nm?). This 1 µm threshold is lower when the cooperative energy transfer between Ytterbium and Erbium is not efficient enough. Phosphorous doping is thus required to ensure short lifetime of the Erbium $^4I_{11/2}$ and avoid back transfer to Ytterbium. Otherwise, this parasitic lasing effect can lead to gain clamping and power limitations for 1.5µm signal generation in Erbium-Ytterbium amplifiers and lasers.

The energy of the Stark sublevels of Yb were computed from the absorption and fluorescence measurements. Several works showed that Ytterbium lasers efficiency were quite sensitive to the fiber temperature [6, 7]. The absorption beyond 1030 nm is mainly caused by the transition b$\rightarrow$ e (1030 nm) whose probability increases with the b level population following the Boltzman statistics. At room temperature, $k_BT \sim 200$ cm$^{-1}$. The population of level b is thus only 7% (fig. 1). At 400 K it grows up to 13%. The absorption in the long wavelength tail of Ytterbium can thus almost double when the fiber temperature increases by 100 K (fig. 2). The Erbium lasing transitions takes place between levels $^4I_{15/2}$ and $^4I_{13/2}$ which are split into 8 and 7 Stark sublevels respectively. The average Stark sublevel separation in each manifold is about 60 cm$^{-1}$. The population is thus almost evenly distributed at room temperature and Erbium cross-section are less sensitive to temperature variations than Ytterbium cross-sections.

Modeling of Er/Yb codoped lasers without temperature dependence has been described elsewhere [8-10]. The cross-relaxation transfer rates were taken from [11]. We have implemented such a numeric model for lasers and amplifier study including computation of the fiber temperature due to the quantum defect and the effect of temperature rising on the initial cross-sections. The power flow in the fiber is first computed using cross-sections at room temperature. The fraction of power dissipated $P_{diss}$ by heating the fiber equals the absorbed pump power minus the signal powers generated at 1µm and 1.5µm (est-ce que ça inclut la fluo, qui est émise isotropiquement? Elle est probablement négligeable, mais on peut la mentionner). We assume that the heat flow is uniform along the fiber. The minimum heat flow dissipated Q in a given fiber cross-section is then estimated using

$$Q = \frac{P_{diss}}{L \cdot \pi a^2} \qquad (1)$$

where a is the core radius, L the fiber length. The fiber temperature can then be computed using a simple cylindrical model with the following equations [3]

$$T_{core} = T_{air} + \frac{Qa^2}{2k}\left(\ln\frac{b}{a} + \frac{k}{bh} + \frac{1}{2}\right) \qquad (2a)$$

$$T_{clad} = T_{air} + \frac{Qa^2}{2bh} \qquad (2b)$$

where a is the fiber core radius, b is the fiber cladding radius and $h = 0.5\, N_u\, \kappa_A\, b^{-1}$ is the air convective parameter related to Nusselt number Nu computed from the Prandtl number Pr and Grasselt number Gr [4]. The thermal conductivity $k = 1.38\ W^{-1}\ m^{-1}$ is assumed to be the same for both glass and outer cladding. $\kappa_A$= 0.026 W m$^{-1}$ K$^{-1}$ is the air conductivity. This is a conservative hypothesis since the outer cladding is usually prepared using a low index polymeric material with very low thermal conductivity.

When the fiber temperature increases, the high energy sublevels in each manifold (b-d and f-g) gets populated following the Boltzman statistics (fig. 1). The absorption beyond 1030 nm is mainly caused by the transition b $\rightarrow$ e (1030 nm) whose probability increases with the b level population. (déjà dit plus haut) The emission beyond 1030 nm is mainly caused by the transitions e $\rightarrow$ b (1030 nm), e $\rightarrow$ c (1087 nm) and f $\rightarrow$ d (1052 nm). However, the transition f $\rightarrow$ d is much weaker than e $\rightarrow$ b as the population of level e is only 4% at room temperature (fig. 1) Dernière remarque pas claire, ou confusion dans les "noms" de niveaux,. We therefore assume that the emission in this region will evolve with temperature as the occupation factor of level e given by $1/Z_u$ were $Z_u$ is the upper partition function.

$$\sigma_e(\lambda,T) = \sigma_e(\lambda,T_{air}) \frac{Z_u(T_{air})}{Z_u(T)} \qquad (4)$$

To compute the evolution of the absorption when the temperature changes we use the Mac-Cumber relation which links the absorption and the emission (fig. 2) [4].

$$\sigma_a = \sigma_e \frac{Z_u}{Z_l} \exp \frac{E_{ZL} - h\upsilon}{k_B T} \qquad (5)$$

The zeroline energy $E_{zl}$ = 10256 cm-1, the upper $Z_u$ and lower partition functions $Z_l$ were computed using the Stark energy levels above. In order to scale the Er cross-sections we used the data from [12] which gives generalized Mac Cumber relations.

We applied this modeling to an Er/Yb laser at 1.55 µm similar to the 103 W laser reported by the Southampton group [1]. Parasitic lasing effects has been observed at the wavelength of maximum Yb gain: 90 W were generated at 1.06 µm versus 103 W at 1.56 µm. Up to 350 W pump power at 975 nm were launched. This high pump power enters the domain where temperature effects can play a significant role. We assumed that the fiber had a 25 µm core and a 400 µm cladding, 60 dB/m Er core absorption, 6 dB/m Yb cladding absorption. The fiber reflectivity was 3.6% at at the pump end (perpendicularly cleaved) and −50 dB at pump wavelength (angle cleaved), 99% above 1520 nm (dichroic mirror) at the other end.

(mes remarques sur la figure 3 sont peut être fausses, car je ne vois pas les symboles sur la figure…) We can observe two kinds of behavior depending on whether the temperature dependence of cross-sections is taken into account or not (fig. 3). Without any temperature dependence, in the small pump rate region, the 1.5 µm power increases steadily. Above about 120 W of pump power, Yb lasing starts at 1060 nm and its population inversion is clamped. It follows that the Erbium population inversion is clamped too and the 1.5 µm output power is quickly limited to about 45 W (sur la figure je devine ~30W (?)). With temperature dependence, the fiber warms up due to pump absorption (figure 4a). For the cross-section scaling we assumed a uniform temperature profile along the fiber. A more detailed power balance allows from the cross-sections previously scaled to compute the local heat generation and temperature profile (fig. 4b???). (je n'ai pas vu d'indication concernant l'extrémité, gauche ou droite, par où la pompe est injectée). The temperature is maximum closed to the pump end with a slight decrease at the pump end due to strong 1um signal generation which reduces the amount of power dissipated. The hypothesis of temperature uniformity across the fiber tends to be inaccurate at very high pump rate. The fiber heating increases the absorption in the Ytterbium tail (fig. 2). The 1 µm lasing threshold is thus increased and the calculation agrees well with the experimental data with the exception of pump powers above 250 W. The discrepancy beyond 250 W may be due to the different approximations done. Suppression of the 3.6 % reflection at pump wavelength on one fiber facet would strongly reduce the 1 µm lasing and 1.5 µm roll-off.

In conclusion, we have demonstrated the influence of thermal effects in high power Erbium-Ytterbium double cladding fiber laser. Due to the pump fraction dissipated, the 1 µm co lasing threshold is increased (ou est-ce son efficacité (pente) qui a décru?) and the 1,5 µm roll-off is delayed (ce verbe ne me semble pas approprié: "shifted toward higher pump rates" serait mieux). Amazingly heat generation can lead to an increased output efficiency in the case of a Erbium-Ytterbium laser showing colasing at 1um and 1.5um. We underline that thermal modeling is thus required in 1.5 µm high power laser with output power larger than 100 W.

Financial support for this research was partially provided by the SOFIA project from region Ile de France. We thank Pascal Dupriez and Christophe Codemard from ORC of University of Southampton for interesting discussions.

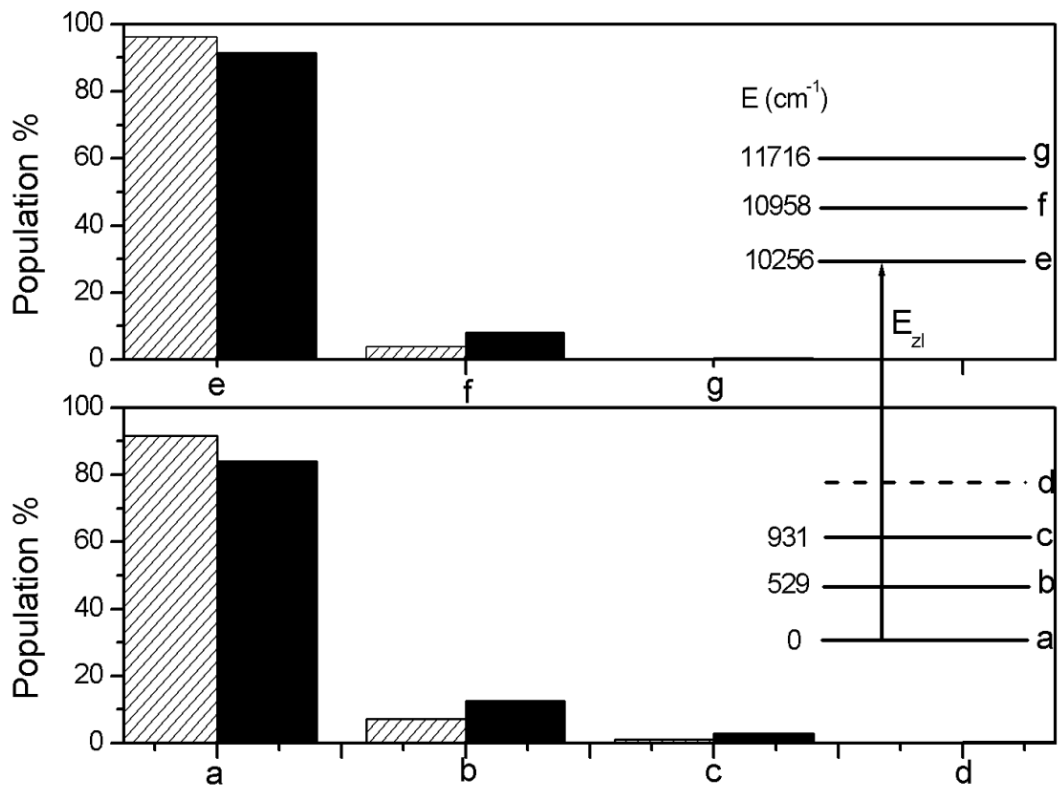

Figure 1: Relative population of the upper (e-g) and lower (a-d) Stark sublevels of Ytterbium (see figure 1) for T=300 K (striped bars) and T=400 K (full boxes). The level structures from absorption and fluorescence measurements is shown on the right.

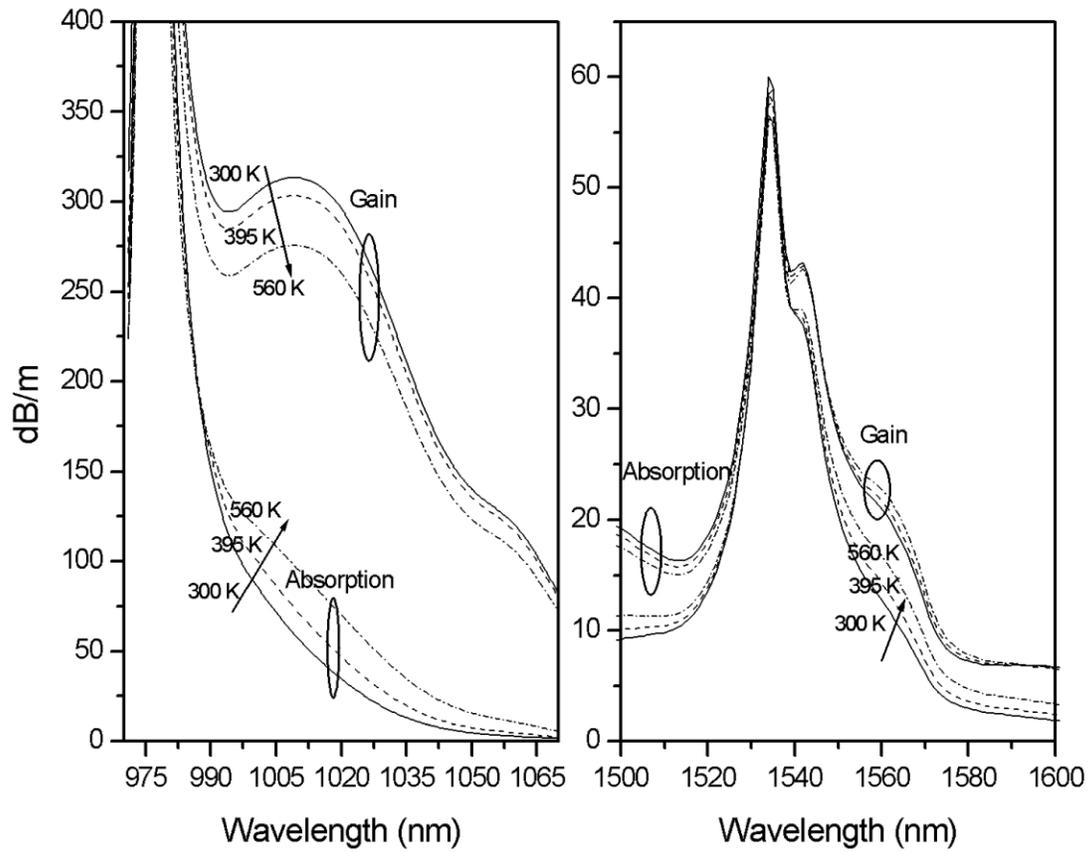

Figure 2: Evolution of the Ytterbium (left) and Erbium (right) absorption and gain with fiber temperature.

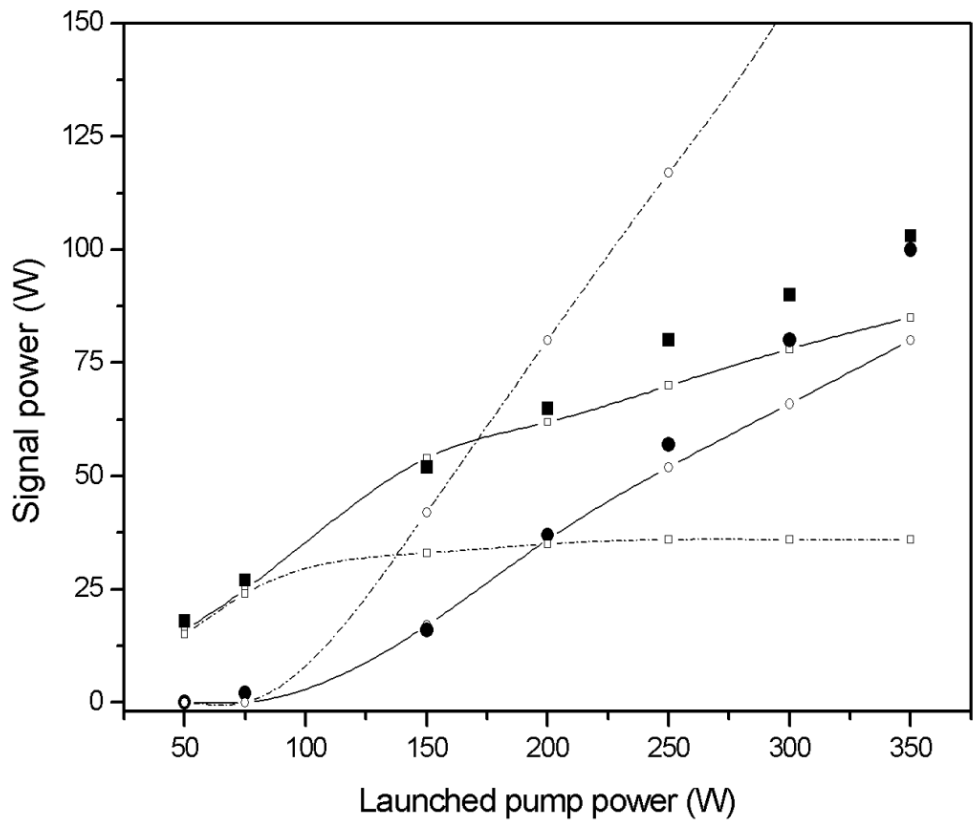

Figure 3: Comparison between experimental values (from [1]) at 1.5 µm (■) and 1 µm (●) ; modeling results with temperature dependence at 1.5 µm (-□-) and 1 µm (-○-) ; modeling results without temperature dependence at 1.5 µm (.-□-.) and 1µm (.-○-.).

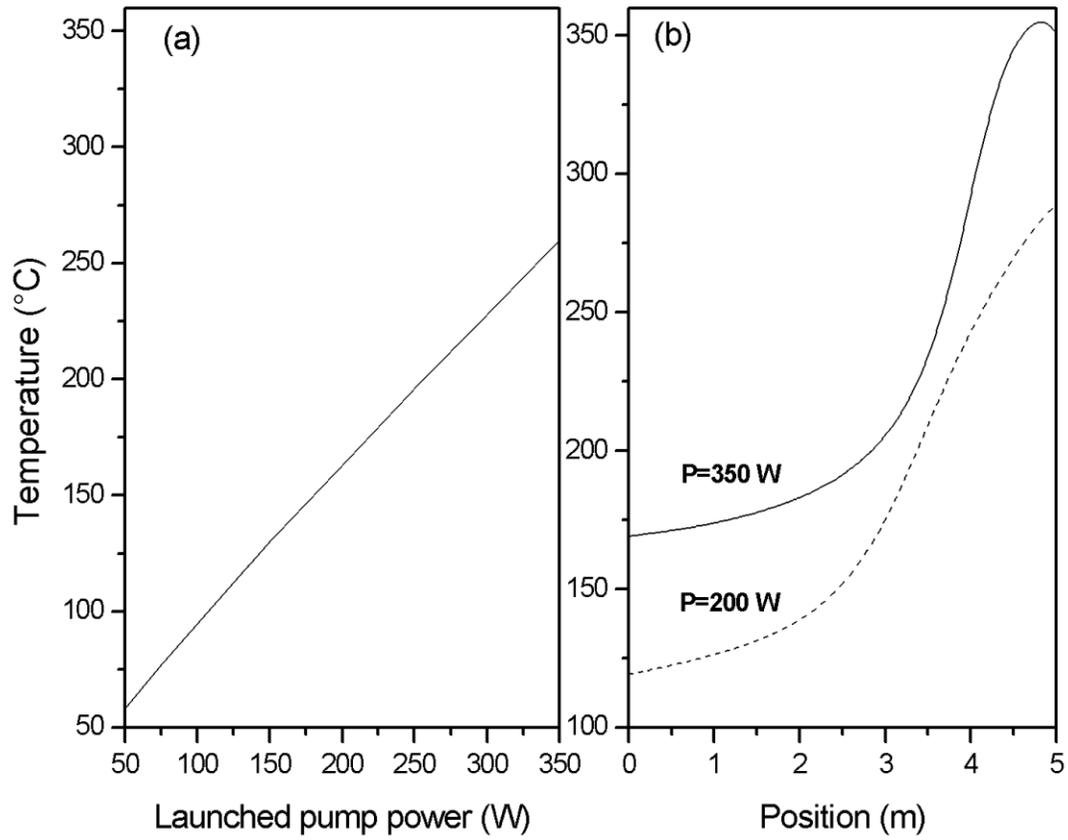

Figure 4: (a): evolution of the fiber temperature with the launched pump power assuming a simplified uniform distribution along the fiber. (b) temperature distribution along the fiber for 350 W pump power (full line) and 200 W pump power (dashed line).

Revoir les légendes, et correspondance avec les figures.

Figure 1: Absorption (full line) and emission (dashed line) cross-sections of Ytterbium used in the modeling. Stark level structure computed from the cross-section showed in the inset.

Figure 2: Relative population of the upper (e-g) and lower (a-d) Stark sublevels of Ytterbium (see figure 1) for T=300 K (striped bars) and T=400 K (full boxes).

Figure 3: Evolution of the Ytterbium (left) and Erbium (right) absorption and gain with fiber temperature.

Figure 4: Comparison between experimental values (from [1]) at 1.5 µm (■) and 1 µm (●) ; modeling results with temperature dependence at 1.5 µm (-□-) and 1 µm (-○-) ; modeling results without temperature dependence at 1.5 µm (.-□-.) and 1µm (.-○-.).

Figure 5: (a) temperature distribution along the fiber for 350 W pump power (full line) and 200 W pump power (dashed line). (b): evolution of the fiber temperature with the launched pump power assuming a simplified uniform distribution along the fiber.